\documentclass{optica-article}

\journal{opticajournal} % for journals or Optica Open

\articletype{Research Article}

\usepackage{lineno}
\usepackage[normalem]{ulem}
\begin{document}

\title{Focal-plane wavefront sensing with moderately broadband light using a short multi-mode fiber}

\author{Auxiliadora Padrón-Brito,\authormark{1,2,*} Natalia Arteaga-Marrero,\authormark{1,2} Ian Cunnyngham,\authormark{3} and Jeff Kuhn\authormark{1,2,3}}

\address{\authormark{1}Instituto de Astrof\'isica de Canarias (IAC), Calle V\'ia Láctea s/n, E-38200 La Laguna, Tenerife, Spain\\
\authormark{2}Departamento de Astrof\'isica, Universidad de La Laguna (ULL), E-38206 La Laguna, Tenerife, Spain\\
\authormark{3}University of Hawai’i at Manoa Institute for Astronomy, 34 'Ōhi’a Street, Makawao, HI 96768, USA}

\email{\authormark{*}maria.auxiliadora.padron@iac.es} %% email address is required; see note below about the corresponding author designation

% use {asbstract*} to suppress the copyright line. Copyright information will be added in production

\begin{abstract*} 
We propose a focal-plane wavefront sensor (FPWFS) based on a short multimode fiber (MMF) capable of operating under moderately broadband illumination. By coupling the aberrated focal-plane field into an MMF of length $\lesssim$1 cm, we preserve modal interference over a 10 nm bandwidth at near-infrared wavelengths. The resulting output intensity pattern encodes pupil phase information, enabling wavefront recovery via a neural network.
Our approach resolves the inherent sign ambiguity of even pupil-phase aberrations and operates on millisecond timescales using readily available computing hardware, suitable for real-time adaptive optics. 
Unlike traditional pupil-plane sensors, the proposed FPWFS shares the optical path with the science beam, eliminating non-common-path aberrations by enabling simultaneous wavefront and focal-plane intensity retrieval. Its simplicity, compactness, sensitivity, and low cost make it an attractive candidate for next-generation astronomical instruments.
\end{abstract*}

%%%%%%%%%%%%%%%%%%%%%%%%%%  body  %%%%%%%%%%%%%%%%%%%%%%%%%%
\section{Introduction}
Light from distant celestial objects arrives at Earth as an almost perfect plane wave.
However, atmospheric turbulence—in the form of rapidly varying refractive‐index inhomogeneities—introduces temporally varying phase distortions.
Adaptive optics (AO) systems attempt to correct these aberrations in real time: a wavefront sensor (WFS) measures the distorted wavefront, and a deformable mirror reshapes itself to cancel out the distortions, restoring an approximation of the original plane wave \cite{davies_adaptive_2012,hampson_adaptive_2021}.

Traditional WFS, such as Shack–Hartmann \cite{shack1971production, platt2001history} or Pyramid sensors \cite{Ragazzoni1996,ragazzoni2002pyramid}, measure phase distortions at the pupil plane. 
Because they share a different optical path than the science camera, they are susceptible to non‐common‐path aberrations (NCPAs). This limits the ability of extreme AO systems to reach the stability required for high-contrast imaging at small angular separations, critical for exoplanet imaging \cite{martinez_speckle_2012,guyon_extreme_2018, savransky2012focalplane}.
They are also vulnerable to low-wind or “petal” effects—phase steps introduced by discontinuities between segments in large, segmented primary mirrors \cite{ndiaye_calibration_2018}.
Similar limitations arise in free-space optical communication (FSOC), where accurate correction of atmospheric turbulence over small angular scales is required to maintain high coupling efficiency into the receiver aperture \cite{wang2018performance}.While FSOC systems typically use a separate wavefront-sensing channel, a focal-plane approach could potentially simplify optical design and reduce NCPAs.

Focal-plane wavefront sensing (FPWFS) circumvents many of these issues by using the same optical path as the science beam \cite{jovanovic2018review_focalplane,savransky2012focalplane}. 
However, for telescopes with a single centrosymmetric aperture and real transmission, reversing the sign of even phase distributions in the pupil—such as defocus or astigmatism—produces the same intensity pattern in the focal plane. %, owing to the symmetry of the Fourier transform for such phase distributions
Iterative algorithms \cite{korkiakoski_fast_2014,bos2020fastfurious, bottom2023} and phase diversity techniques \cite{Kendrick1994phasediversity,keller2012extremely} can lift this degeneracy, but require multiple exposures or splitting the beam, consuming observing time, and often involve significant computational effort.
Machine learning- phase retrieval \cite{Andersen19,orban2021focal,quesnel2022deep,wang2024use,bottom2023} reduces the computational burden, yet still needs to resolve the sign ambiguity either through using defocused images or phase masks like a vortex coronagraph.
While compatible with exoplanet imaging, where coronagraphs are already employed, this requirement reduces general applicability and prevents use in scenarios like FSOC.
An alternative strategy involves non-mode-selective photonic lanterns, which adiabatically transform a multimode input into an array of single-mode outputs, enabling wavefront encoding in the single-mode intensities \cite{norris_all-photonic_2020,lin2023focal,lin_focal-plane_2022,wei2024photoniclantern}.

\begin{figure}[h]
    \centering
    \includegraphics[width=1.0\textwidth]{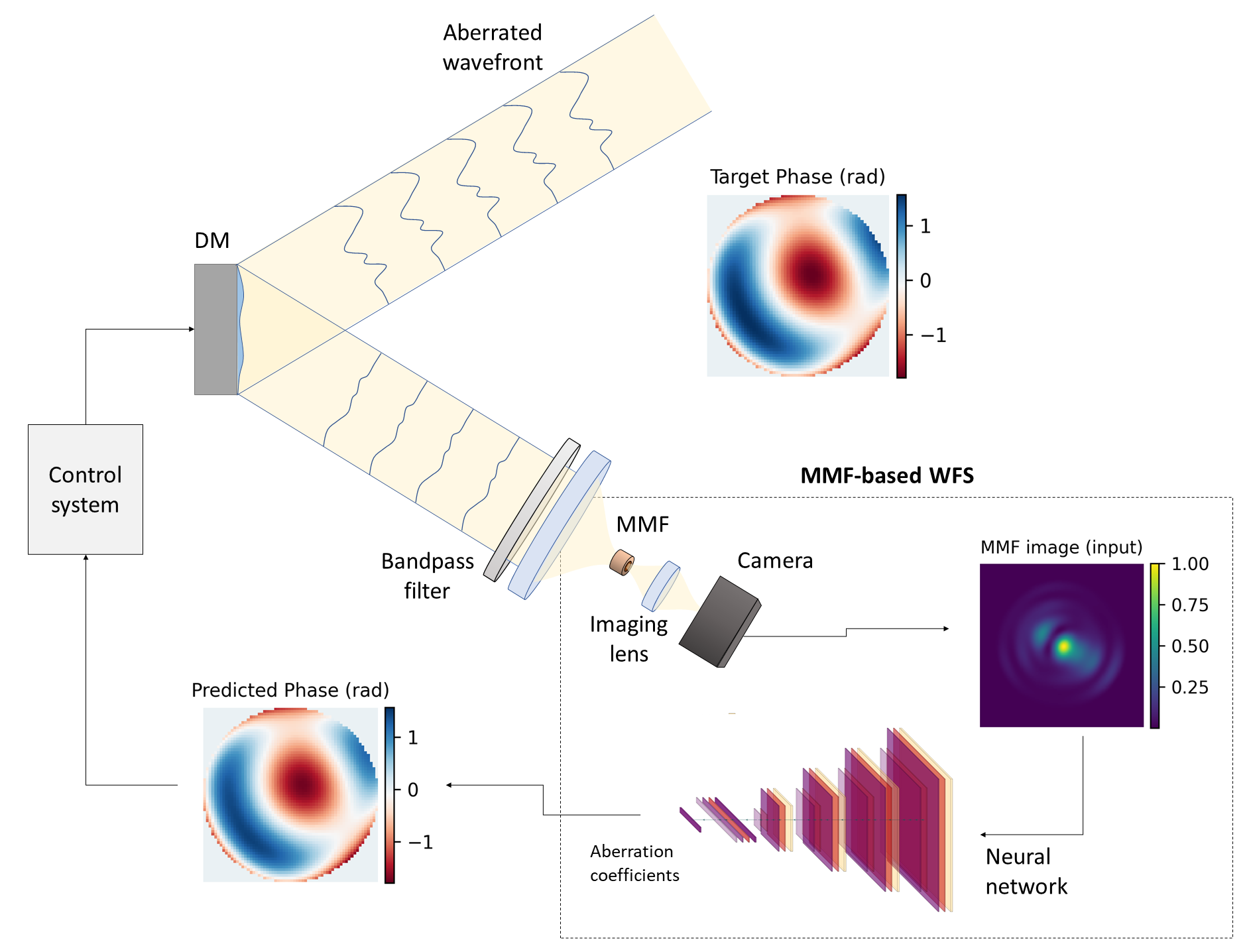}
    \caption{Concept of the MMF‑based FPWFS in a closed‑loop AO system. An aberrated wavefront enters the telescope and is partially corrected by a deformable mirror (DM). The residual wavefront is spectrally filtered and coupled into a short MMF. The emerging intensity pattern is captured by a camera and processed by a NN, which retrieves the pupil‑plane phase and inform the control system. This sends a feedback signal back to the DM. An example pair of target and NN‑predicted wavefronts is shown.}
    \label{fig:MMF_WFS_concept}
\end{figure}

We propose a simpler approach: the distorted broadband focal-plane field is coupled into a single short multimode fiber (MMF), and the resulting output intensity pattern is imaged onto a camera. A neural network (NN) extracts the pupil-plane phase information from the measured intensity and provides feedback to the AO deformable mirror. The concept is illustrated in Fig. \ref{fig:MMF_WFS_concept}.
Because the same MMF output intensity pattern also encodes the focal-plane intensity—that is, the intensity at the fiber entrance—it is possible to recover the science image by training a second NN. This would constitute a true FPWFS.
Since the effective field of view (FOV) is restricted by the fiber core, this method is compatible with applications where the region of interest is intrinsically narrow, such as high-contrast imaging of Earth-like exoplanets or FSOC.

Under the weak‐guidance approximation, the input field can be decomposed into a set of guided fiber modes (LP modes), each of which accumulates a different phase while propagating through the MMF. 
If the MMF length is short enough that the differences in group delay between modes remain well below the coherence time, interference fringes are generally preserved rather than averaged out.
The resulting output pattern is therefore highly sensitive to both the amplitudes and relative phases of the excited modes—and hence to the full complex input wavefront—making it possible to reconstruct the wavefront purely from the measured intensity.

Recovering the pupil-plane phase from the output intensity pattern of the MMF is a nonlinear problem, which can be solved by training a NN on pairs of known wavefronts and MMF output images \cite{sinha2017lensless}.

Although the use of MMFs with machine‑learning algorithms has been proposed for bioimaging and optical communications \cite{borhani_learning_2018,rahmani_multimode_2018,zhu_image_2021,wang_multimode_2023,caramazza_transmission_2019}, adapting them to astronomical wavefront sensing entails three principal challenges.
First, unlike laboratory settings where a narrowband laser or reference beam can be employed, astronomical observations rely on starlight that is intrinsically broadband and inaccessible for manipulation at the source. Second, the use of extremely narrow spectral filters to enhance coherence length markedly reduces photon throughput—thereby degrading the signal‐to‐noise ratio (SNR). Third, standard‐length MMFs are intrinsically sensitive to mechanical vibrations and thermal drifts in observatory environments, leading to variations in the fiber’s transmission matrix and necessitating frequent recalibration. Nonetheless, NNs for related MMF applications have demonstrated robustness under variable external conditions \cite{wen_single_2023,abdulaziz_robust_2023}.

The idea of using a conventional long MMF as a WFS in astronomy has been previously proposed and preliminarily explored \cite{DiFrancesco2024}, though only accounting for the monochromatic case. To the best of the author's knowledge, this is the first demonstration of its operation under broadband illumination relevant for astronomical application.

In this work, we show, through both simulations and experiments, that a short step-index MMF can preserve interference fringes when its length is appropriately matched to the source bandwidth.
We further show that this property enables the resolution of the sign ambiguity associated with even phase distributions at the pupil.

Building on this, we generate a dataset of simulated input wavefronts and corresponding MMF output intensity patterns. Using this data, we train a convolutional neural network (CNN) to infer the first 11 Zernike modes from the fiber output intensity. We achieve an average root-mean-square error (RMSE) of 0.03 radians in total phase prediction for small input aberrations (input RMSE < 0.6 radians), and an average per-coefficient RMSE of 0.018 radians.
Once trained, the CNN delivers wavefront predictions on millisecond timescales with standard processing hardware, fast enough to follow atmospheric fluctuation.
This approach yields promising results, indicating the potential of this method for real-time aberration sensing in astronomical AO systems.

\section{Materials and Methods}

\subsection{Experimental setup}
In order to study the effect of the spectral bandwidth in the MMF output intensity pattern, we used two different sources: a laser diode with a wavelength of 1064 nm, and a quartz halogen light source. The latter was coupled to a single-mode fiber to clean the spatial mode and spectrally filtered using a bandpass filter with a center wavelength of 1 $\mu$m and a bandwidth of 10 nm.

\begin{figure}[h]
    \centering
    \includegraphics[width=0.5\textwidth]{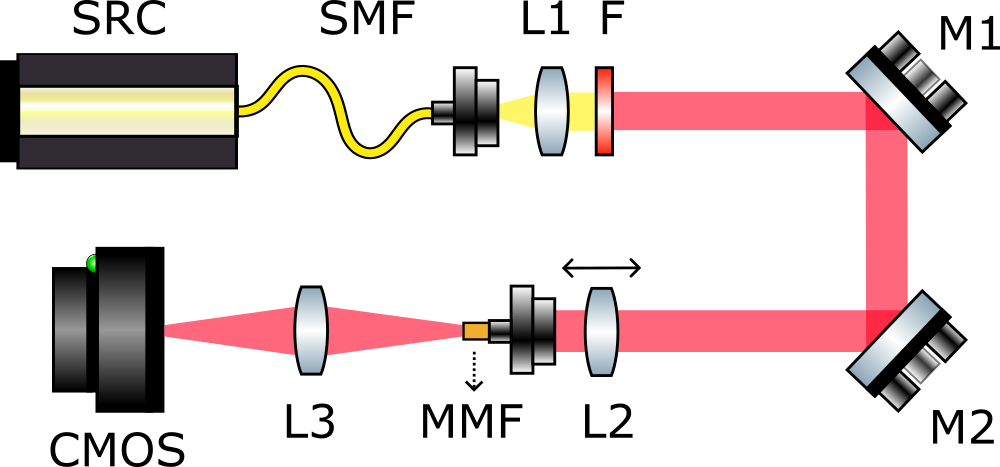}
    \caption{Schematic of the general experimental setup. SRC: source, SMF: single-mode fiber, L1: collimating lens, F: spectral filter (used only with the white source), M1 and M2: mirrors, L2: movable focusing lens, L3: imaging lens, CMOS: imaging camera.}
    \label{fig:setup}
\end{figure}

A picture of the setup can be seen in Figure \ref{fig:setup}. Light coming from the source is collimated using a fiber collimator package (L1) with a focal length of 11.17 mm and $\text{NA}=0.25$. 
The transmitted light is in-coupled into a MMF by using an aspheric lens (L2) with focal length of 18.4 mm.

We used custom‑cut MMFs with a core radius $r_{\text{core}}=25$ $\mu$m and numerical aperture $\text{NA}=0.22\pm0.02$. 
These parameters are well suited for coupling low-order aberrations, as $r_{\text{core}}$ is slightly larger than the Gaussian beam waist at the fiber input ($\sim5$ $\mu$m).

For a telescope system, $r_{\text{core}}>r_{\text{Airy}
}$ and $\text{NA}>\text{NA}_{\text{tel}}$ 
ensure that the fiber core fully captures the central lobe of the diffraction-limited point-spread function (PSF) and the entire cone of the focal-plane point-source field used in the AO system. In the presence of aberrations, the fiber core should remain larger than the aberrated PSF to maintain efficient coupling.

If the proposed MMF-based FPWFS is also used to recover the focal-plane image, the effective FOV is limited by the fiber core size by $\text{FOV}\sim 2r_{\text{core}}/f$.

Two fiber lengths were prepared—one measured as $9.92\pm0.01$
mm, and another approximately 1 m long—to probe the effects of modal delay with different bandwidth sources. In all configurations, the coupling efficiency exceeded 93\% under optimal alignment. 

The output of the MMF is then imaged by an optical system formed by an achromatic lens (L3) and a CMOS camera.

In order to test the distinguishability of even phase distributions in the pupil, we generated defocus by translating the in‑coupling lens (L2) along the optical axis.
This adjustment must be performed with care to avoid inadvertently adding odd‑order aberrations such as tip/tilt, which could be detected directly in the focal‑plane image.

\subsection{Simulations}
We simulated the light propagation and coupling to the MMF using the HCIPy package\cite{por2018hcipy}, a comprehensive toolkit for high-contrast imaging simulations in Python.

In our simulations, we modeled a circular pupil with a diameter of 3 mm and an optical system with an effective focal length of 18.4 mm, corresponding to an approximately $f/6$ system with NA$_{\text{tel}}\approx D/(2f)\approx0.08$. The MMF was modeled as a step-index fiber with a core radius $r_{\text{core}}=25$ $\mu$m, numerical aperture $\text{NA}=0.22$, and variable length \textit{L}. These parameters were selected to replicate the experimental conditions, acknowledging that they differ from those of an actual telescope system.
At a wavelength $\lambda=1$ $\mu$m, the diffraction‐limited Airy radius is $r_{\text{Airy}}=1.22\lambda f /D \approx$ 7.5 $\mu$m, much smaller than the fiber core radius, which ensures efficient coupling of a slightly aberrated field.

The pupil phase distribution $\phi$ was expressed as a linear combination of the first 11 Zernike polynomials. This number should be enough to account for the most common aberration types present in the experimental setup. Moreover, atmospheric distortions \cite{Fried:66, Noll:76}, as well as NCPAs are dominated by the lowest-order Zernike modes.

The electric field at the pupil plane for a given wavelength was defined as $E_{\text{pupil}} = \text{Aperture} \cdot e^{i\phi}$, where \textit{Aperture} was 1 inside the pupil and 0 outside.
The focal-plane electric field was obtained by applying a Fraunhofer propagator. This field was then propagated through the MMF by calculating its LP modes and their propagation constants $\beta$. For the chosen fiber parameters, there were 597 guided modes.

Using the fiber mode basis, a transformation matrix between the fiber modes and the focal-plane grid was constructed. The input electric field was then projected onto the fiber modes to determine the excitation coefficients. After propagation along the fiber length $L$, each mode accumulated a phase factor $e^{i\beta L}$. Finally, the output electric field was transformed back to the focal-plane grid basis.

To simulate finite-bandwidth light propagation, we summed the intensities resulting from a sufficiently dense sampling of wavelengths.
To determine an appropriate wavelength step size, we employed an empirical method: we progressively decreased the spacing between adjacent wavelengths until the output speckle patterns from consecutive wavelengths became visually indistinguishable. We found that direct visual comparison was more sensitive to subtle pattern changes than global metrics such as RMSE. For a MMF length of 9.92 mm, convergence was achieved with a $\sim$0.05 nm step, resulting in 201 propagated wavelengths for the full 10 nm bandwidth.

\subsection{Machine learning}

CNNs are particularly well-suited for solving imaging problems, as they extract spatial features from images through successive convolutional and pooling layers \cite{fukushima1982neocognitron,lecun1998gradient}. Then, they offer a natural framework for inferring input phase information from complex output intensity images \cite{sinha2017lensless}. 
Accordingly, CNNs have been widely adopted in MMF-related applications in imaging and optical communications, where they are used to reconstruct or classify images transmitted through the fiber \cite{borhani_learning_2018, rahmani_multimode_2018, zhu_image_2021}. 
Most of these works rely on established architectures such as VGG \cite{simonyan2014very} or U-Net \cite{ronneberger2015u}.
However, alternative approaches have also been explored \cite{zhu_image_2021, caramazza_transmission_2019}.

In this work, a supervised CNN was trained to predict the pupil Zernike coefficients—excluding the piston term, since WFS are inherently insensitive to a global phase offset—from the intensity pattern at the MMF output. The dataset compised 82,000 simulated samples, with 80\% used for training, 20\% for validation (5-fold cross-validation), and an additional 4,000 samples reserved for testing. Zernike coefficients were sampled from a normal distribution with zero mean and a standard deviation of $\pi/16$ rad, leading to aberrations well below the phase-wrapping limit and within the system’s sensitive range. This choice was not based on a fundamental limitation; larger aberrations could be explored in future work.

\begin{figure}[h]
    \centering    \includegraphics[width=0.95\textwidth]{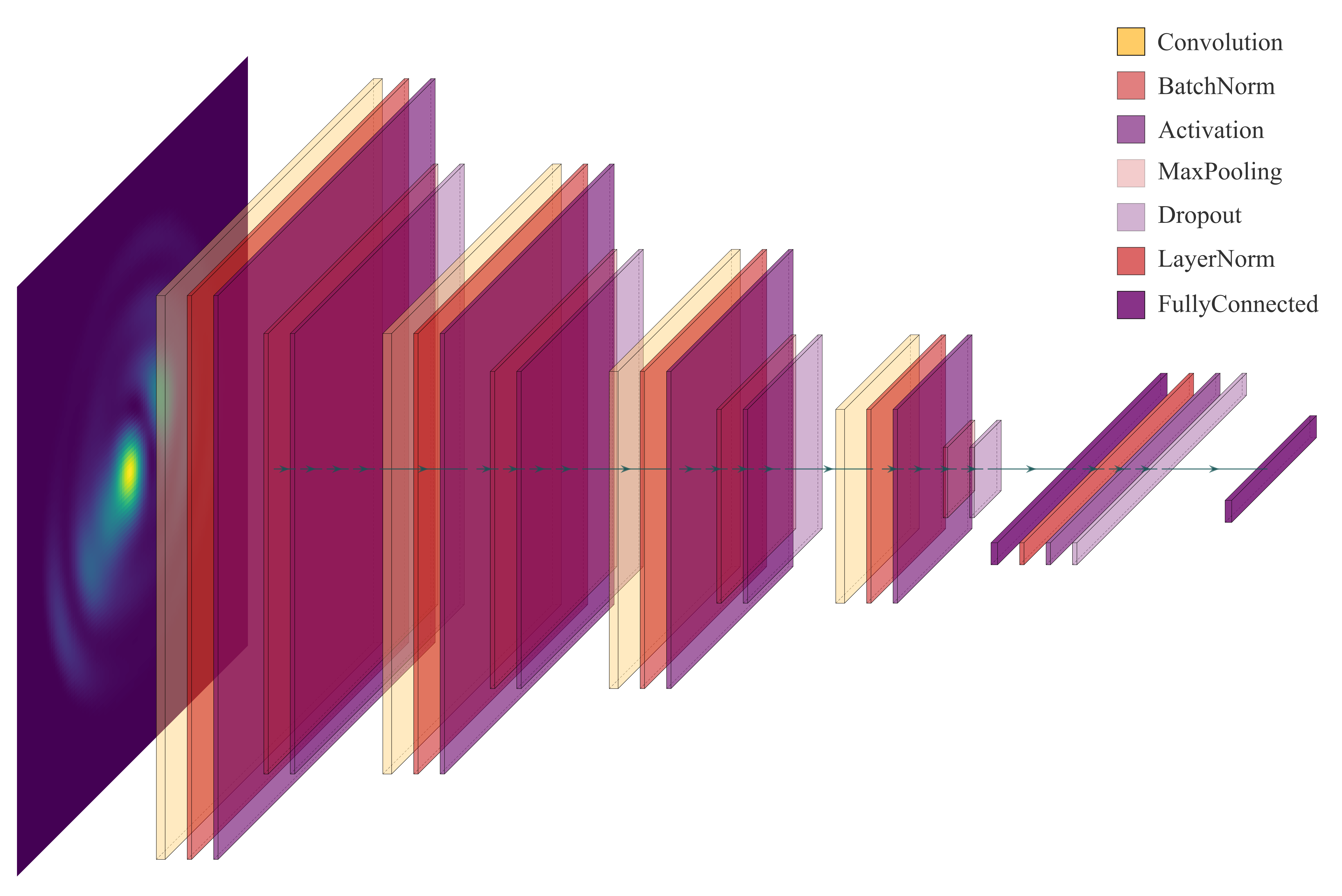}
    \caption{Schematic of the selected CNN architecture \cite{iqbal2018plotneuralnet}. Note that the spatial dimensions of the subsequent layers are not drawn to scale and serve for illustrative purposes only.}
    \label{fig:my_arch_4}
\end{figure}

The network consisted of four convolutional layers with batch normalization, tanh activations, max-pooling, and 25\% dropout, followed by two fully connected layers. The first included layer normalization, ReLU activation and 25\% dropout, whereas the second is used to predict the coefficients. An scheme of the CNN architecture is shown in Figure \ref{fig:my_arch_4}. The model was intentionally kept shallow to prioritize speed while maintaining sufficient accuracy.

Mean Squared Error (MSE) was used as the loss function, and the network was trained with the Adam optimizer (learning rate of 0.001), early stopping (patience of 10 epochs), and model checkpointing. Training used a batch size of 16 for 26 epochs with a fixed random seed for reproducibility.
Notice that the CNN was trained on a system equipped with a 2 x Intel Xeon 6354 CPUs (each 18 cores, 2 threads, 3.00 GHz) and an NVIDIA A100-SXM4 GPU with 80 GB of memory. For inference, an Intel Core i9-10900X CPU (10 cores, 20 threads, 3.70 GHz) paired with an NVIDIA Quadro RTX 4000 GPU (8 GB) was used.

\section{Results and Discussion}

\subsection{Break of sign degeneracy of even phases}

We showed that interference between modes in a MMF, arising from their different propagation constants, can break the sign ambiguity of even pupil-phase distributions (see Supplementary Material for the theoretical background). For a broadband source, this modal interference is preserved only if the differential group delay between modes is shorter than the source coherence time. This sets a \emph{coherence-limited maximum fiber length}  
\begin{equation}\label{eq:fiber_length}
z_{\max} \lesssim \frac{\lambda^2}{\Delta\lambda\,\Delta n_{g,max}},
\end{equation}
where $\Delta n_{g,max}$ is the maximum group-index difference between modes (of order $n_{\mathrm{core}}-n_{\mathrm{clad}}$ for weakly guiding step-index fibers).

\subsubsection*{Simulations}

To illustrate the resolution of the sign ambiguity, we selected defocus as a representative even Zernike polynomial. The conclusions drawn from this example extend to other even modes.

\begin{figure}[h]
    \centering
    \includegraphics[width=0.95\textwidth]{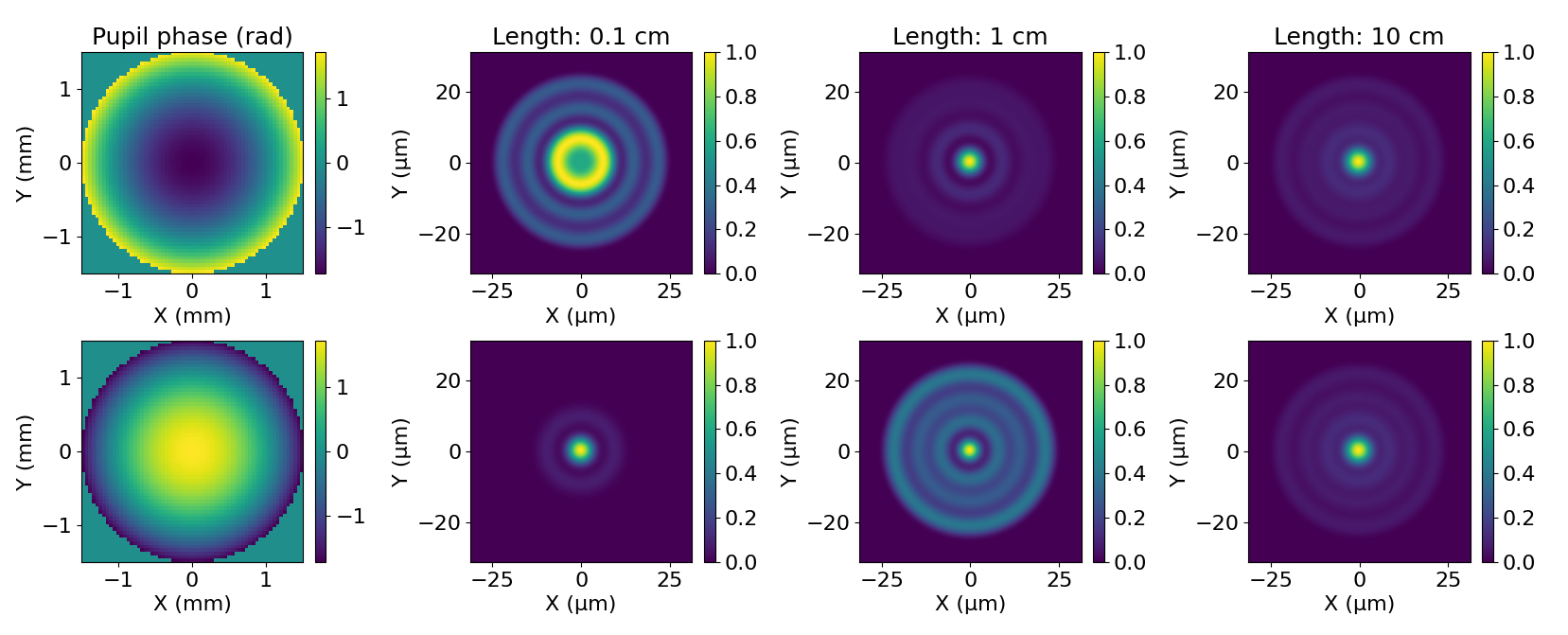}
    \caption{Comparison of normalized intensity at the output of a MMF for inverted defocus (top and bottom row; corresponding pupil phases shown on the left), for a light source with central wavelength of 1 $\mu$m and bandwidth $\Delta \lambda = 10$ nm, with varying fiber lengths. The intensity patterns from the 10 cm fiber barely distinguish the input sign difference.}
    \label{fig:astig_broad}
\end{figure}

The effect of a finite-bandwidth source is shown in Fig.~\ref{fig:astig_broad}, where we compare the MMF output patterns for different fiber lengths using a 10 nm bandwidth centered at $\lambda = 1 \, \mu$m. The intensity patterns emerging from the MMF were normalized to its maximum value. For 0.1 cm and 1 cm-long MMFs, the structure of the output intensity is still sensitive to the sign of the input phase. However, for a 10 cm-long MMF, the output patterns become nearly indistinguishable. The \emph{coherence-limited maximum fiber length} inferred from Eq. \ref{eq:fiber_length} is on the order of $z_{max}\sim1$~cm, which is in agreement with the simulated data.

\subsubsection*{Laboratory test}
\begin{figure}[h!]
    \centering
    \includegraphics[width=0.99\textwidth]{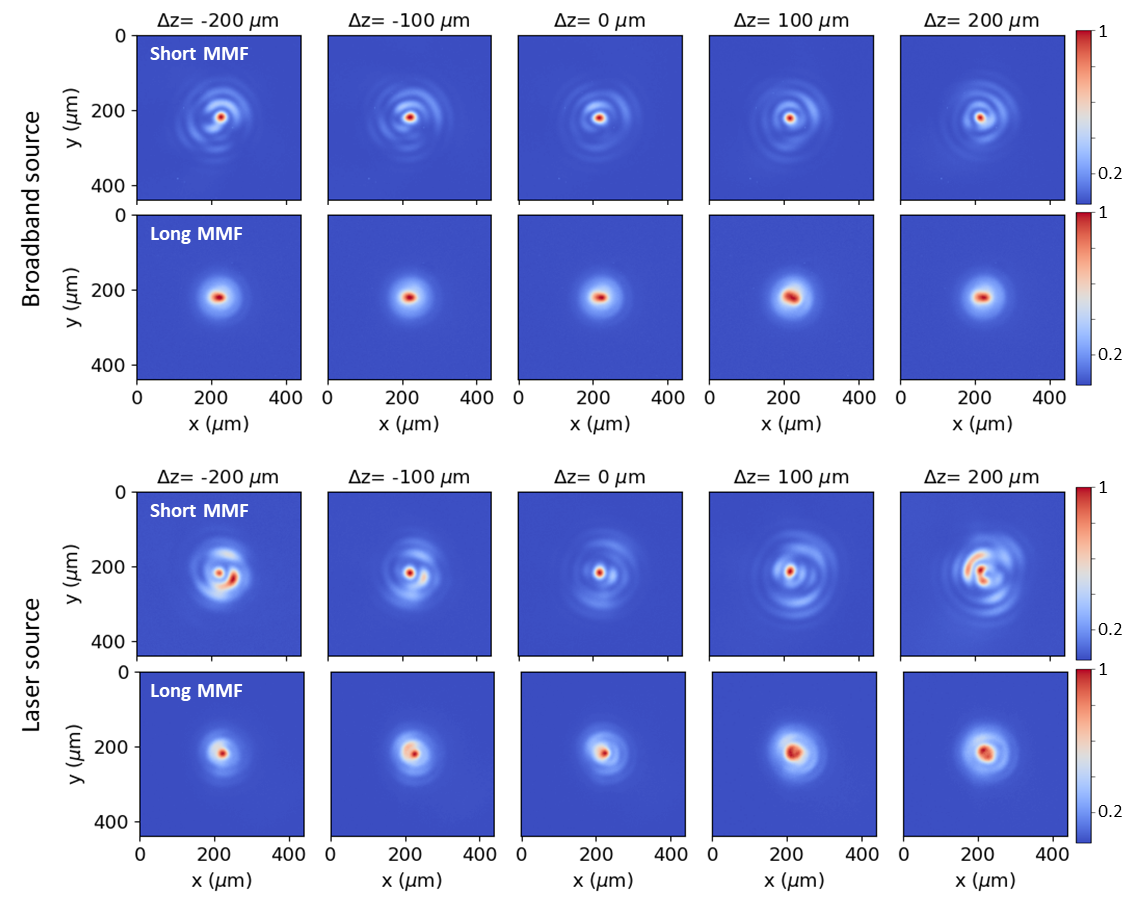}
    \caption{Impact of fiber length and source bandwidth on the output image of the MMF. Top panel: source with 10 nm bandwidth. Bottom panel: 1064 nm laser. In each case, the top row shows the output image from a 9.92 mm-long MMF, while the bottom row corresponds to a $\sim1$ m-long MMF. Left to right: output images measured for different positions of the in-coupling lens along the optical axis. $\Delta z = 0$ denotes the position of optimized coupling.}
    \label{fig:exp_data}
\end{figure}

We validated our simulation results with a laboratory test. To generate an even phase distribution resembling defocus at the pupil, we shifted the in-coupling lens (L2 in Fig.~\ref{fig:setup}) slightly forward and backward from its optimal position, along the optical axis. The resulting intensity patterns at the output of a MMF were recorded under four different conditions, varying both the fiber length and the spectral bandwidth of the input light, as shown in Fig.~\ref{fig:exp_data}.

The upper panel of Fig.~\ref{fig:exp_data} shows results obtained with the broadband white-light source, filtered by a 10~nm band-pass filter centered at 1~$\mu$m. The first row corresponds to the 9.92~mm-long MMF output, and the second row to a $\sim$1~m-long fiber. The lower panel presents analogous measurements for the laser source at 1064~nm, with the third and fourth rows corresponding to the short and long fiber, respectively.

In these measurements, the output patterns for the short fiber (first and third rows) appeared richer in spatial structure than those for the long fiber (second and fourth rows), suggesting stronger modal interference in the short fiber case. Under the short-fiber condition, the patterns also changed noticeably when the in-coupling lens was displaced toward positive or negative defocus. This asymmetry was consistent with the MMF retaining some sensitivity to the sign of even phase distributions, although noise and other aberrations may have influenced the results.

In contrast, for the case of the 10~nm bandwith and long fiber (second row), the output intensity was nearly uniform across the core, with little variation as the lens position changed. This observation was consistent with a loss of sensitivity to the sign of the input phase when the modal delay exceeded the light’s coherence length ($z_{max} \sim 1$~cm). % For a 10~nm bandwidth, Eq.~\ref{eq:fiber_length} predicts a maximum useful fiber length on the order of 1~cm, in agreement with the experimental trend.

\subsection{Phase reconstruction via machine learning}

To evaluate the system's ability to recover pupil-plane phase distributions from MMF output images, we employed a CNN trained on simulated data. The training dataset consisted of paired samples of 11 Zernike coefficients, which defined the input pupil phase, and the corresponding intensity patterns at the MMF output.

\begin{figure}[h!]
    \centering
    \includegraphics[width=0.9\textwidth]{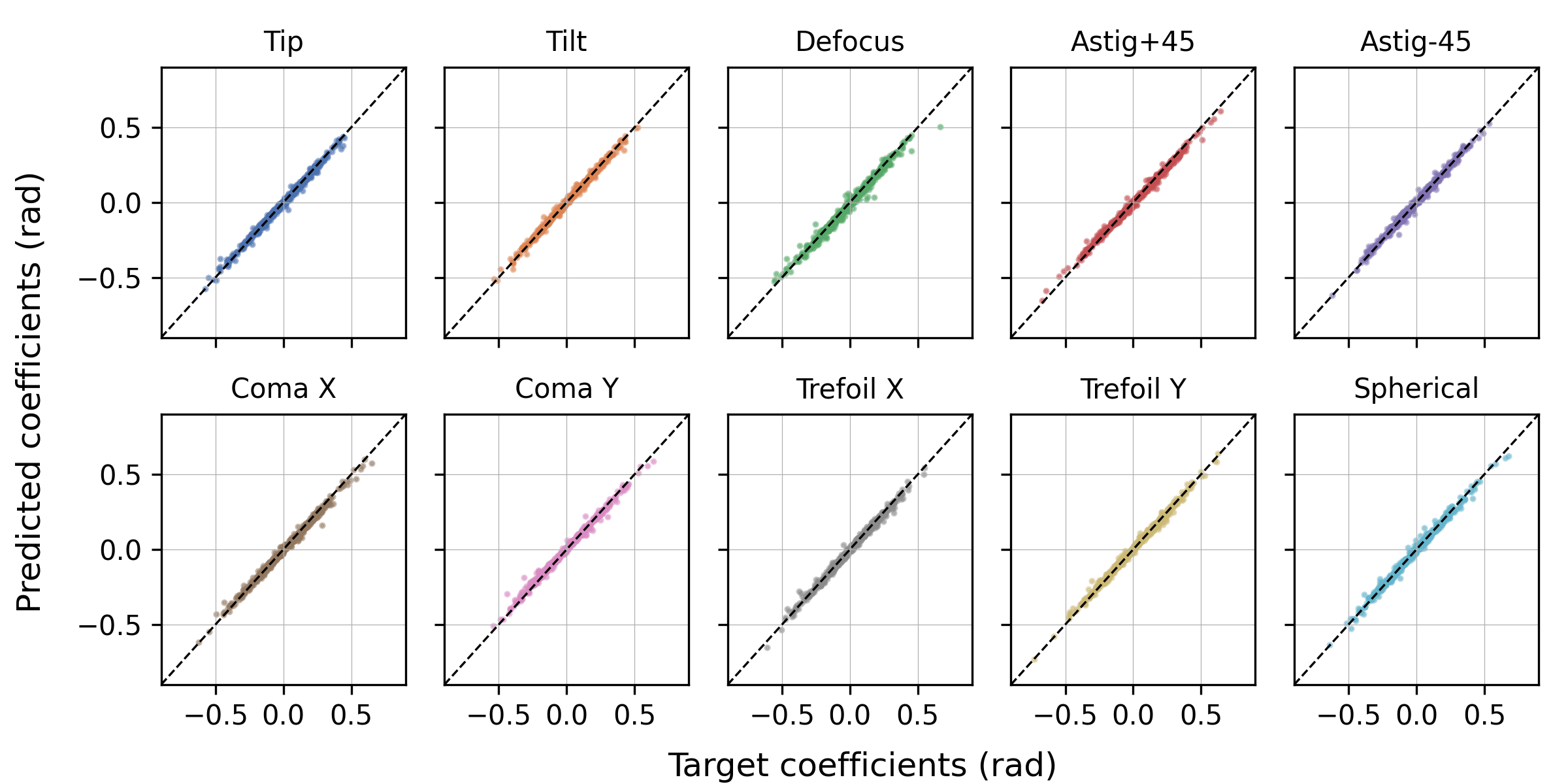}
    \caption{Predicted versus target values of Zernike coefficients (in radians), for the first 11 Zernike polynomials, excluding piston. $y=x$ line is depicted as a reference in dashed black line.}
    \label{fig:Pred_vs_target_coeff}
\end{figure}

In Fig.~\ref{fig:Pred_vs_target_coeff}, we show a scatter plot comparing the predicted and target values of each Zernike coefficient (in radians). The predictions align closely with the identity line ($y = x$), indicating a strong agreement between model output (prediction) and ground truth (target).

\begin{table}[htbp]
\caption{Performance of the CNN for each Zernike coefficient, excluding piston. 
RMSE (rad) is the root mean squared difference, $R^2$ is the coefficient of determination between predicted and target values, and Bias (rad) is the mean difference.}
    \centering
    \begin{tabular}{cccc}
        \hline
        \textbf{Coefficient} & \textbf{RMSE (rad)} & \textbf{R$^2$} &  \textbf{Bias (rad)}\\
        Tip & 0.018 & 0.991 & -0.0008\\
        Tilt & 0.017 & 0.992 & 0.0022\\
        Defocus & 0.024 & 0.986 & -0.0074\\
        Astigmatism (+45º) & 0.021 & 0.989 & -0.0021\\
        Astigmatism (-45º) & 0.017 & 0.992 & 0.0031\\
        Coma X & 0.018 & 0.991 & -0.0039\\
        Coma Y & 0.020 & 0.990 & 0.0057\\
        Trefoil X & 0.017 & 0.993 & -0.0044\\
        Trefoil Y & 0.017 & 0.992 & 0.0033\\
        Spherical & 0.018 & 0.992 & 0.0006\\
        \hline
    \end{tabular}
    \label{table:Metrics_coefficients}
\end{table}

Table~\ref{table:Metrics_coefficients} reports the prediction performance for each Zernike coefficient using three standard metrics: RMSE, coefficient of determination ($R^2$), and bias. Across all coefficients, RMSEs remain below 0.025~rad, $R^2$ values exceed 0.98, and biases are close to zero, indicating that the CNN provides accurate and unbiased estimates of the target coefficients. The error does not increase with the order of the Zernike coefficients, suggesting that the method could accommodate more complex phase structures. %Although a study of how the number of Zernike polynomials affects the error is beyond the scope of this work, it will be addressed in future studies.

\begin{figure}[h!]
    \centering    \includegraphics[width=0.85\textwidth]{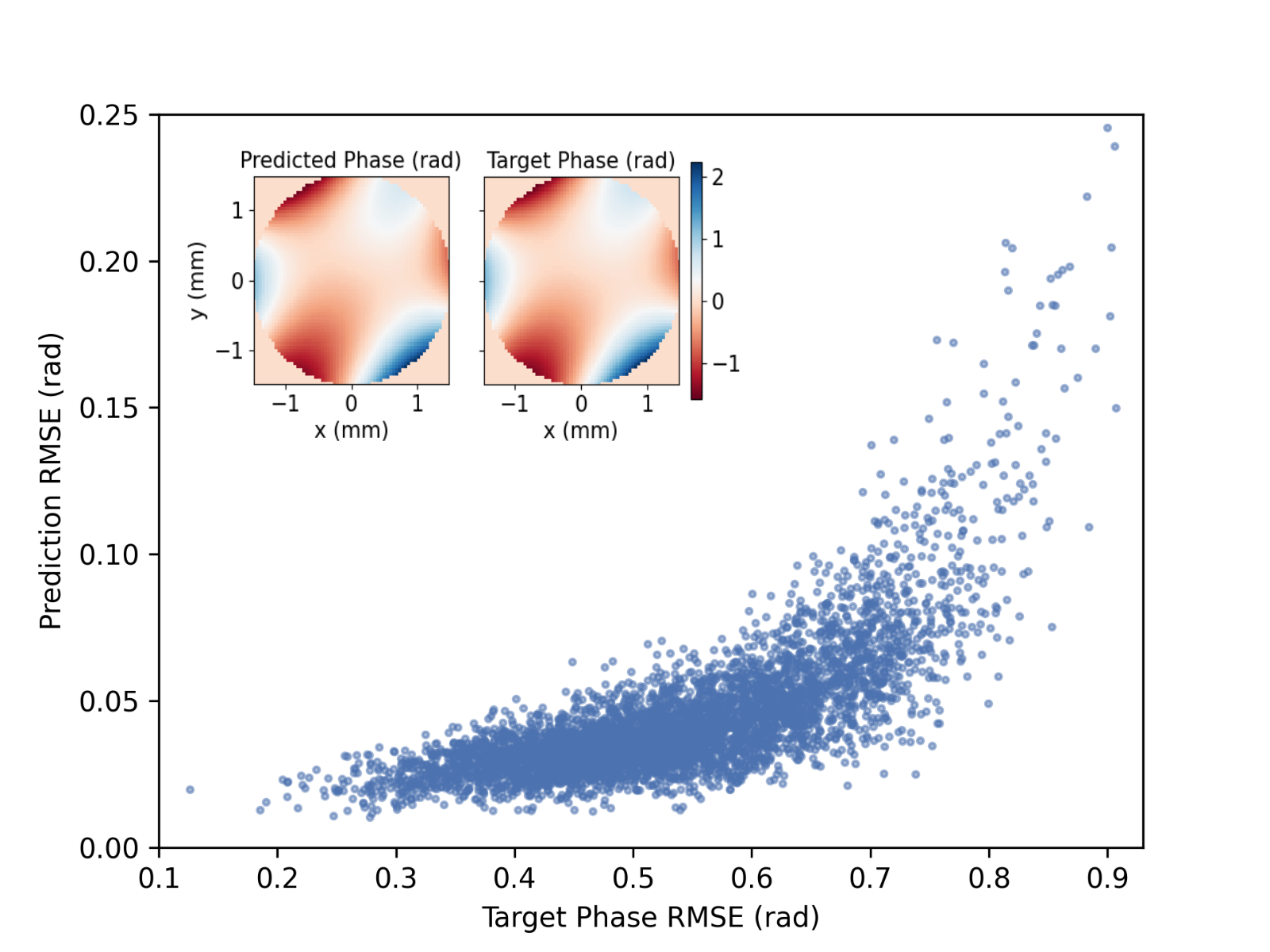}
    \caption{Prediction RMSE versus Target Phase RMSE (in radians). The prediction RMSE quantifies the model’s accuracy, computed as the RMSE between the phase reconstructed from the predicted Zernike coefficients and that from the target coefficients. An example of the reconstructed phases is plotted in the inset. The target phase RMSE reflects the level of aberration in the input data, measured as the RMSE between the reconstructed target phase and its mean value.}
    \label{fig:accuracy_prediction_RMSE}
\end{figure}

The prediction accuracy of the CNN as a function of input aberration is shown in Fig.~\ref{fig:accuracy_prediction_RMSE}. To generate this plot, we first reconstructed the full pupil-phase maps from the Zernike coefficients (an example is plotted in the inset of Fig.~\ref{fig:accuracy_prediction_RMSE}). We then computed two metrics for each input: the \textit{prediction RMSE}, defined as the pixel-wise root-mean-square difference between the predicted and target phase maps, which quantifies the overall phase prediction error; and the \textit{target phase RMSE}, defined as the root-mean-square deviation of the target phase relative to its mean, which serves as an indicator of the input aberration strength.

As observed in Fig.~\ref{fig:accuracy_prediction_RMSE}, the \textit{prediction RMSE} increases with the level of input aberration. For inputs with \textit{target phase RMSE} below 0.6~rad, the average \textit{prediction RMSE} is 0.034~rad, while for weakly aberrated inputs (\textit{target phase RMSE} $< 0.3$~rad), the average \textit{prediction RMSE} decreases to 0.022~rad.  

This trend might arise because stronger aberrations at the pupil generate more complex and highly variable intensity patterns at the MMF output. An ill-conditioned mapping from output intensity to input wavefront, combined with the sparsity of high-dimensional data, might cause larger prediction errors that grow with aberration strength. Additionally, the MMF parameters (core radius and NA) were chosen to optimize sensitivity to weak aberrations. Larger pupil aberrations could reduce the fraction of light efficiently coupled into and transmitted through the fiber, further limiting the sensor’s sensitivity to strong aberrations.

The CNN architecture achieves prediction times consistently below 1.5~ms per sample, using a conventional system for inference, well within the coherence time of atmospheric turbulence.

Table~\ref{table:Metrics_training_size} illustrates the influence of training dataset size on CNN performance. As expected, larger datasets improve both the coefficient prediction accuracy and the reconstructed pupil-phase RMSE, though diminishing returns are observed beyond $\sim 10^5$ samples.

\begin{table}[htbp]
\caption{Performance of the CNN for different training dataset size. \textit{Coeff. RMSE} (rad) is the average RMSE between predicted and target Zernike coefficients. \textit{Phase RMSE} (rad) is the average RMSE between the reconstructed pupil-phase maps from the predicted and target coefficients.}
    \centering
    \begin{tabular}{cccc}
        \hline
        \textbf{Dataset size} & \textbf{Training time (h)} & \textbf{Coeff. RMSE (rad)} &  \textbf{Phase RMSE (rad)}\\
        164000 & 7.43 & 0.017 $\pm$ 0.006 & 0.042 $\pm$ 0.027\\
        82000 & 4.68 & 0.018 $\pm$ 0.007 & 0.045 $\pm$ 0.028\\
        16400 & 1.50 & 0.021 $\pm$ 0.008 & 0.053 $\pm$ 0.034\\
        8200 & 1.12 & 0.026 $\pm$ 0.010 & 0.067 $\pm$ 0.041\\
        1640 & 0.05 & 0.14 $\pm$ 0.04 & 0.39 $\pm$ 0.12\\
        \hline
    \end{tabular}
    \label{table:Metrics_training_size}
\end{table}

\section{Conclusion}

We showed that a compact MMF could serve as a low-cost FPWFS capable of resolving the sign ambiguity of even pupil-phase aberrations under moderately broadband illumination. Using a CNN, we achieved phase-reconstruction errors as low as 22~mrad for residual input aberrations below 0.3~rad, with prediction times under 1.5~ms per sample— consistent with real-time extreme AO requirements. While these results demonstrate the feasibility of the approach, laboratory and on-sky demonstrations are needed to assess its robustness under noise sources and environmental variations.

Beyond step-index fibers, graded-index MMFs could extend the permissible fiber length by reducing modal dispersion. More generally, alternative waveguide geometries—such as laser-written waveguides in thin glass substrates—could  provide compact, flexible implementations.

The MMF output intensity encodes both focal-plane images and pupil-plane wavefronts, enabling the design of NCPA-free sensors for applications that demand strong wavefront correction over very narrow fields of view, such as exoplanet imaging. In FSOC, including quantum links, a practical approach could employ a narrowband beacon channel to meet the coherence and SNR requirements of the MMF-based sensor.

In summary, combining a short MMF with fast CNN inversion provides a compact, low-mass, and low-cost focal-plane alternative to conventional WFSs, while being particularly sensitive to small aberrations. Unlike iterative or phase-diversity FPWFS, it avoids photon losses and heavy computation, and in comparison to other ML-based approaches, it resolves even-phase ambiguities without requiring defocused images or coronagraphs. Its simplicity makes it competitive with photonic lantern–based FPWFS, while remaining compatible with integrated photonic platforms and suitable for next-generation AO instruments.

\begin{backmatter}
\bmsection{Funding}
The authors acknowledge funding from the European Union, Project Ref.: 101087032. Views and opinions expressed are however those of the authors only and do not necessarily reflect those of the EU or the EU Research Executive Agency. Neither the European Union nor the granting authority can be held responsible for them. LIOM project's R\&D\&i activities are also supported by the Cabildo Insular de Tenerife thanks to the ``Apoyo a las actuaciones I+D+I en el espacio de cooperación IACTEC" collaboration agreement.
\bmsection{Acknowledgment}
The authors thank Félix Gracia Témich for custom‑cutting the MMFs used in the experiments, and the team at the Laboratory for Innovation in Optomechanics (LIOM) for useful discussions.
The authors acknowledge Light Bridges for sharing computing resources, particularly the ICR-Astronomy$\textsuperscript{\textregistered}$ infrastructure.
\bmsection{Disclosures} All authors are listed as inventors on a patent related to the technology described in this work. The authors declare no other conflicts of interest.
\bmsection{Data availability} Data underlying the results presented in this paper are not publicly available at this time but may be obtained from the authors upon reasonable request.
\end{backmatter}

%%%%%%%%%%%%%%%%%%%%%%% References %%%%%%%%%%%%%%%%%%%%%%%%%

%%%%%%%%%% If using BibTeX:
\bibliography{References}

\newpage
\section{Supplementary material}
\subsection{Break of sign degeneracy of even phases: monochromatic case}

For a monochromatic input, the pupil-plane electric field can be written as
\begin{equation}
u_{\mathrm{in}}(x,y) = t(x,y)\, e^{\,i\phi(x,y)}
= t(x,y)\,\cos\phi(x,y) + i\,t(x,y)\,\sin\phi(x,y),
\end{equation}
where $t(x,y)$ is the real-valued pupil transmission, and $\phi(x,y)$ is the phase aberration.

Let $\mathcal{F}$ denote the (unitary) discrete Fourier transform. We define
\begin{equation}
a(\xi,\eta) = \mathcal{F}\!\left\{ t(x,y) \cos\phi(x,y) \right\}, 
\quad
b(\xi,\eta) = \mathcal{F}\!\left\{ t(x,y) \sin\phi(x,y) \right\}.
\end{equation}
Here, $(\xi,\eta)$ denote the transverse Cartesian coordinates in the focal-plane spatial-frequency domain, conjugate to the pupil-plane coordinates $(x,y)$. Therefore, the focal-plane field is
\begin{equation}
u_{\mathrm{foc}}(\xi,\eta) \propto a(\xi,\eta) + i\,b(\xi,\eta).
\end{equation}
The proportionality symbol is used here to omit the overall phase and scaling constants in the Fraunhofer propagator.

If $t(x,y)$ and $\phi(x,y)$ are both even functions, then $t\cos\phi$ and $t\sin\phi$ are also real and even, and hence $a(\xi,\eta)$ and $b(\xi,\eta)$ are real (and even).

Let us project the focal-plane field onto the $n$ LP modes of the multimode fiber (MMF). Let $\psi_j(\xi,\eta)$ denote the (real-valued) transverse profiles of the LP modes, which satisfy the orthonormality condition
\begin{equation}
\iint \psi_j(\xi,\eta)\,\psi_k(\xi,\eta) \,d\xi\,d\eta = \delta_{jk}.
\end{equation}
The modal coefficients at the MMF input ($z=0$) are
\begin{equation}
C_j(0) = \iint \big[a(\xi,\eta) + i\,b(\xi,\eta)\big]\, \psi_j(\xi,\eta)\,d\xi\,d\eta,
\qquad j=1,\dots,n.
\end{equation}
Writing the real and imaginary projections explicitly,
\begin{equation}
A_j := \iint a(\xi,\eta)\,\psi_j(\xi,\eta)\,d\xi\,d\eta,
\qquad
B_j := \iint b(\xi,\eta)\,\psi_j(\xi,\eta)\,d\xi\,d\eta,
\end{equation}
we have
\begin{equation}
C_j(0) = A_j + i\,B_j, \qquad j=1,\dots,n.
\end{equation}

Propagation over a fiber length $z$ multiplies each modal coefficient by a phase factor:
\begin{equation}
C_j(z) = e^{i\beta_j z}\, C_j(0) = e^{i\beta_j z}\,\big(A_j + i\,B_j\big),
\end{equation}
where $\beta_j$ is the propagation constant of mode $j$.

The field at the output of the MMF is then
\begin{equation}
u_{\mathrm{out}}(\xi,\eta) = \sum_{j=1}^n e^{i\beta_j z}\,
\big[ A_j + i\,B_j \big]\; \psi_j(\xi,\eta).
\end{equation}

Replacing $\phi \to -\phi$ changes the sign of the $\sin\phi$ term:
\begin{equation}
(a,b) \;\longrightarrow\; (a,-b)
\quad\Rightarrow\quad
(A_j,B_j) \;\longrightarrow\; (A_j,-B_j),
\end{equation}
so that
\begin{equation}
u_{\mathrm{out}}^{(-\phi)}(\xi,\eta) =
\sum_{j=1}^n e^{i\beta_j z}\,
\big[ A_j - i\,B_j \big]\; \psi_j(\xi,\eta).
\end{equation}
If all $e^{i\beta_j z}$ are equal (up to a global phase), $u_{\mathrm{out}}^{(-\phi)}$ is the complex conjugate of $u_{\mathrm{out}}^{(\phi)}$ and their intensities are identical—i.e., the sign degeneracy remains. Moreover, if one measures \emph{modal} intensities (e.g., via a mode-selective photonic lantern), then
\begin{equation}
|C_j(z)|^2 = |C_j(0)|^2 = A_j^2 + B_j^2,
\end{equation}
which is also independent of the sign of $\phi$. See also \cite{lin_focal-plane_2022}.

When the $\beta_j$ differ, the fiber introduces mode-dependent phase shifts. This prevents $u_{\mathrm{out}}^{(-\phi)}$ from being a pure conjugate of $u_{\mathrm{out}}^{(\phi)}$, and their output intensities differ. This is the degeneracy-breaking mechanism.

Expanding the output intensity, with $\Delta\beta_{jk} := \beta_j - \beta_k$ and using that $\psi_j$ are real, we obtain
\begin{equation}\label{eq:int}
\begin{aligned}
\big| u_{\mathrm{out}}^{(\pm \phi)} (\xi,\eta)\big|^2
&= \sum_{j=1}^n \big( A_j^2 + B_j^2 \big)\, \psi_j^2(\xi,\eta) \\
&\quad + 2\sum_{j<k}\!\left[ \big(A_jA_k+B_jB_k\big)\,\cos(\Delta \beta_{jk} z)
\mp \big(B_jA_k - A_jB_k\big)\,\sin(\Delta\beta_{jk} z)\right]\,\psi_j(\xi,\eta)\,\psi_k(\xi,\eta).
\end{aligned}
\end{equation}
Thus, only the $\sin(\Delta\beta_{jk} z)$ cross term flips sign under $\phi\to-\phi$ (for even $\phi$), providing the observable signature that breaks the sign degeneracy.

\subsection{Broadband case}
For the case of a broadband source, one should take into account that the propagation constants are wavelength-dependent, $\beta_j(\lambda)$. The relative phase accumulated between modes is then also wavelength-dependent, $\Delta \beta_{jk}(\lambda) z$. For a broadband source coupled to a sufficiently long fiber, these wavelength-dependent modal phases cause the interference terms in the output intensity (second term of Eq.~\ref{eq:int}) to wash out, leaving a nearly incoherent sum of the individual modal intensities (first term of Eq.~\ref{eq:int}).

To formalize this, we should take into account the \emph{modal dispersion}. Different modes propagate at different group velocities $v_{g,j} = c / n_{g,j}$, where $n_{g,j}$ is the group index of mode $j$.
The differential propagation time between two modes $j$ and $k$ over a fiber of length $z$ is:
\begin{equation}
\Delta t_{jk}(z) = z \left( \frac{1}{v_{g,j}} - \frac{1}{v_{g,k}} \right) 
= \frac{z}{c} \, \Delta n_{g,jk},
\end{equation}
with $\Delta n_{g,jk} = n_{g,j} - n_{g,k}$. For a source with coherence time $\tau_c$, modal interference is preserved only if $\Delta t_{jk} \lesssim \tau_c$. For a narrowband Gaussian-like source with a central wavelength $\lambda$ and bandwidth $\Delta\lambda$, the coherence time is approximately
\begin{equation}
\tau_c \approx \frac{\lambda^2}{c \, \Delta \lambda}.
\end{equation}
Combining these relations, and taking $\Delta n_{g,max}$ to be the \emph{maximum} group index difference between any two modes (for weakly guiding step-index fibers, $\Delta n_{g}$ is on the order of $n_{\mathrm{core}} - n_{\mathrm{clad}}$), we obtain an upper bound on the fiber length for which modal interference—and thus degeneracy breaking—can still be observed:
\begin{equation}\label{eq:zmax}
z_{\max} \sim \frac{\lambda^2}{\Delta\lambda \, \Delta n_{g,max}}.
\end{equation}

\subsection{Considerations about the minimum fiber length}
On the other hand, to observe sign-degeneracy breaking, the fiber must be long enough for the $\sin(\Delta\beta_{jk} z)$ term in Eq.~\ref{eq:int} to be significantly nonzero. In the monochromatic case for two modes $j$ and $k$, the first point where sign degeneracy is \emph{maximally} broken occurs when
\begin{equation}
\Delta\beta_{jk} z = \pi \quad\Rightarrow\quad z_{jk,\min} \approx \frac{\pi}{|\Delta\beta_{jk}|}.
\end{equation}
For a weakly guiding, step-index fiber with $\beta_j \in [ 2\pi n_{\mathrm{clad}}/\lambda, \, 2\pi n_{\mathrm{core}}/\lambda ]$, this yields an order-of-magnitude estimate:
\begin{equation}\label{eq:zmin}
z_{\min} \sim \frac{\lambda}{2\,(n_{\mathrm{core}} - n_{\mathrm{clad}})}.
\end{equation}

In summary, broadband sources impose a \emph{maximum} fiber length (coherence-limited) beyond which modal interference washes out, while modal beating imposes a \emph{minimum} length (beating-limited) needed to break the sign degeneracy of even pupil-phase aberrations.

The coexistence of the beating-limited minimum fiber length $z_{\min}$ and the coherence-limited maximum fiber length $z_{\max}$ requires
\begin{equation}
    z_{\min} < z_{\max}.
\end{equation}
Using Eqs.~\eqref{eq:zmin} and \eqref{eq:zmax}, this condition translates into an upper bound on the source bandwidth:
\begin{equation}
\Delta\lambda_{\max} \;\sim\; 2\,\lambda\,\frac{\Delta n_g}{\,n_{\mathrm{core}}-n_{\mathrm{clad}}\,}.
\end{equation}
For weakly guiding step-index fibers, where $\Delta n_g \lesssim n_{\mathrm{core}}-n_{\mathrm{clad}}$, this yields the loose estimate
\begin{equation}
\Delta\lambda_{\max} \;\lesssim\; 2\,\lambda.
\end{equation}
Although this suggests that bandwidths up to $\Delta\lambda_{\max} \sim 2\lambda$ could, in principle, be supported, achieving such a condition requires a fiber length near $z_{\min}$, which is on the order of tens of microns for the parameters of the optical system used in this study. Fibers of this length are impractical to fabricate or handle. Consequently, the usable bandwidth in practice is significantly smaller, and the exact limit depends on the chosen fiber length and its modal dispersion characteristics.

The discussion above assumes that the input field actually excites guided modes of the fiber at the entrance plane ($z=0$). This requires overlap of the field with both core and cladding. If, for a sufficiently short fiber, the field remains almost entirely within the core and does not reach the core–cladding boundary, the propagation can be approximated as that in a homogeneous medium of refractive index $n_{\text{core}}$.

A very coarse estimate of the minimum propagation distance for core–cladding interaction can be obtained from ray optics as $z_{\text{min}}\sim r_{\text{core}}/\text{NA}$, where NA is the input field numerical aperture, and $r_{\text{core}}$ the fiber core radius.

For a Gaussian input beam with waist radius $\omega_0$ at the fiber facet, a more refined estimate is

\begin{equation}
    z_{\text{min}} \approx z_{R} \sqrt{\left(\frac{r_{\text{core}}}{\omega_0}\right)^2-1}
\end{equation}
where $z_R$ is the Rayleigh range defined by $z_R = \frac{\pi \omega_0^2 n_{\text{core}}}{\lambda}$.

For the parameters of the optical system used in this study, $z_{\text{min}} \lesssim 0.5$ mm.

\end{document}